

\font\twelvebf=cmbx10 scaled\magstep 1
\font\twelverm=cmr10 scaled\magstep 1
\font\twelveit=cmti10 scaled\magstep 1

\font\tenbf=cmbx10
\font\tenrm=cmr10
\font\tenit=cmti10

\font\ninerm=cmr9

\parindent=1.5pc
\hsize=6.0truein
\vsize=8.5truein
{ \bf January 1995 \hfill PUPT-1525 \break}
{ \null \hfill hep-th/9501021}
\baselineskip=22pt
\centerline{\bf FIELD THEORY AND}
\baselineskip=16pt
\centerline{\tenbf THE PHENOMENON OF TURBULENCE}
\vglue 0.8cm
\centerline{\tenrm V. Gurarie}
\baselineskip=13pt
\centerline{\tenit Department of Physics, Princeton University}
\baselineskip=12pt
\centerline{\tenit Princeton, NJ 08544, USA}
\vglue 0.8cm
\centerline{\tenrm ABSTRACT}
\vglue 0.3cm
{\rightskip=3pc
 \leftskip=3pc
 \tenrm\baselineskip=12pt\noindent
We study the phenomenon of turbulence from the point of view of
statistical physics. We discuss what makes the turbulent states different
from the thermodynamic equilibrium and give the turbulent analog of
the partition function. Then, using the soluble theory of turbulence of waves
as an example, we construct the turbulent action and show how one can
compute the turbulent correlation functions perturbatively thus developing
the turbulent Feynman diagrams. And at last, we discuss which part
of what we learnt from the turbulence of waves can be used in other
types of turbulence, in particular, the hydrodynamic turbulence of
fluids.
This paper is based on the talk delivered at SMQFT (1993) conference at the
University of Southern California.

\vglue 0.6cm}
\vfil
\twelverm\baselineskip=14pt
\leftline{\twelvebf 1. Introduction}
\vglue 0.3cm
The phenomenon of turbulence has been known to physicists for more than a
century and yet it remains to be one of the unsolved problems of
modern physics. Its formulation is extremely simple. The behavior of
incompressible fluid is governed by two simple equations$^1$
$$ \eqalign { \rho \left ({ \partial {\bf v} \over \partial t}
+ ({\bf v \nabla})
{\bf v} \right) &= - {\rm \bf grad}~p + \nu \Delta {\bf v} \cr
   {\rm div}~{\bf v} = 0} \eqno (1.1)$$
for the unknowns ${\bf v}$ and $p$, the velocity of the fluid at a given
point and
the pressure. For a given problem one just needs to solve these
equations with the appropriate boundary conditions to find the motion
of the fluid completely, as it seems. But as was known for a long
time, these so called laminar solutions describe what we observe
only when the velocity is small enough.
As the velocity increases, the
solutions for the Eq.~(1.1) become unstable and the fluid switches to
a new regime of a very complex motion with the velocity pulsating almost
randomly and without any noticeable order. To describe what exactly
is going on when the fluid is in such regimes is of extreme importance
to both the fundamental physics and the applications.

Since its discovery, the enormous amount of efforts has been put into
the studying of
turbulence$^{1,2}$.  It was realized that the most simple
situation where the turbulence exhibits itself is when we somehow
shake the fluid homogeneously in its entire volume. This is
called the homogeneous turbulence$^{1,2}$ and this is what we are
going to study here.

Of course, the first natural thing to do when approaching this problem
is to realize that the random pulsations of the velocity leave us with
little else to do other than to pass to the statistical description.
Instead of the velocity ${\bf v}$ we should work with the average
$\langle {\bf v} \rangle$. The average here is supposed to be over the time.
We can also define the correlation functions like $\langle {\bf v}({\bf x})
{\bf v}({\bf y}) \rangle$. This is exactly what is done in statistical physics
when studying complex systems with many degrees of freedom. But statistical
physics is a well developed branch of  science. It teaches us that
the averages over the time should be replaced by the averages over the space
with the probability distribution. The probability distribution is
usually found on the one of the first pages of any book on statistical
physics.   It is simply
$$ P = \exp ( - { E \over T} ) \eqno (1.2) $$
where $E$ is the energy of the system, and $T$ is called the temperature.
Eq.~(1.2) is called Gibbs distribution.
And so as long as the energy is known, we are left just with a
mathematical problem of computing the averages of $P$ over the
phase space. For the motion described by Eq.~(1.1) with the absence
of the viscosity $\nu=0$
the energy can
be shown to be$^1$
$$ E=\int { {\bf v}^2 \over 2}  \eqno (1.3) $$
and the integral over the phase space turns out to be a Gaussian path
integral. In fact, the phase space unit volume should also be properly
defined before doing the actual calculations, but all those difficulties
can be overcome and the correlation functions can be computed exactly
(see, for example, the reference [4]).

However, it does not mean we solved the problem of turbulence. The
turbulent motion of fluid is not described by Gibbs distribution.
It is because the motion of fluid is
essentially dissipative. The dissipation is described by the term $\nu \Delta
{\bf v}$ and is very important for the whole phenomenon. While
the energy conserving case $\nu=0$ is properly  described by the
thermodynamic equilibrium distribution given above, switching on
dissipation, even infinitesimally weakly, completely changes the whole
picture. Since in real life all the motion is with dissipation,
the thermodynamic calculations are of little importance to physics of fluids,
apart from a few very special cases like the atmosphere of Jupiter etc.$^4$

What really makes the picture is that the constant motion of fluid
must be supported by some external force which injects the energy
into the fluid. This energy is dissipated with the same rate as it is
injected to provide for some form of a dynamic equilibrium.

In this form the problem again looks almost unapproachable. It looks
as if the correlation functions essentially depend on how exactly
we inject the energy and what kind of stirring force we use to do the job.
But already in twenties it was realized that there was still
some universality left in the phenomenon$^{1,2}$.

The energy consumption depends on $\Delta {\bf v}$ and that means, the smaller
the wavelength, the bigger the dissipation. It may turn out (and it actually
does) that the energy dissipation is unimportant at wavelengths bigger than
a certain small value, while the stirring force acts only at some
large wavelengths characteristic for a given source of energy. In between
there is what is called the inertial range where we may neglect
both the stirring force and the viscosity. The situation here is
reminiscent of that of the phase transition theory. There is some cutoff
(two cutoffs in our case)  and a universal picture
in the middle. In the inertial range the Eq.~(1.1) with $\nu=0$
works, but yet we cannot use the thermodynamic distribution
 because we observe a flux of energy from large to small scales through
the inertial range. All the energy comes to a given wave from a larger
one, and it goes away to smaller ones unlike the thermodynamic equilibrium
where the detailed balance principle is satisfied, that is there is as much
energy transferred from one wave to another as there is energy transferred
back. This picture of energy cascade through the system is characteristic
of turbulence. In fact, when we will be speaking of turbulence in this
paper, we will mean the systems with fluxes of energy or other conserved
quantities. This is what really defines the phenomenon of turbulence
and makes it so difficult to study.

There exist some statistical states which are characterized by the
fluxes which are just small deviations from the thermodynamic equilibrium.
These are studied in the kinetics. However, they have nothing to do with
turbulence. The turbulent states are far away from the thermodynamic
equilibrium and the notion of temperature has no sense for them$^{6}$.

 Kolmogorov was the first to attempt a quantitative study
of this picture. He said that
the only relevant parameter which could enter any correlation function
was
the energy transfer rate per unit mass
$\epsilon$ (if we express everything in terms of the unit mass,
then the density of fluid $\rho$ should not enter any formula).
Then some of the correlation
functions can immediately be found by a simple dimensional analysis.
For example, let us find the energy
distribution over different wavelengths  $E(k)$. Its dimension is
${ cm^2 \over s^2  } cm$ because $E(k) dk$ must have the dimension
of the energy per unit mass.
$\epsilon$ is the energy per unit mass per unit time, so its
 dimension is ${ cm^2 \over s^2} {1 \over  s}$. By comparing the
dimensitons, we immediately obtain
$$ E(k)={1 \over 2} \langle {\bf v} (k) {\bf v}(-k) \rangle \propto
\epsilon^{2 \over 3} k^{-{5 \over 3}} \eqno (1.4)$$
This is called the Kolmogorov spectrum. Its coordinate version can
be written as
$$  \langle ({\bf v} ({\bf x})- {\bf v}(0))^2 \rangle \propto x^{2 \over 3}
\eqno (1.5)$$
which implies the dimension of the velocity to be $-{1 \over 3}$.
The Kolmogorov spectrum was measured in many experiments, numerical and
``real-life" and was found to be very close to $-{5 \over 3}$ although
there is no  particular agreement on the exact value of that number.
The proximity of the measured value of
the spectrum to the predicted one has persuaded many researchers that
$-{ 5 \over 3}$ is in fact an {\twelveit exact} value and many efforts had
been spent in the attempts to prove it. And yet, even now, 50 years after
Eq.~(1.4) was first written, we do not know if it is exact or not.
The reason for us to doubt the Eq.~(1.4) is that the considerations
which lead us to it are essentially mean field theory.
We now know very well from the study of critical phenomena that
mean field theory often fails$^5$ because the cutoffs may enter
explicitly the formulae we are trying to find. That is why
we are still in need of constructing  a quantitative theory
of turbulence not based on any conjectures.

Most of the attempts of proving or disproving Eq.~(1.4) were based on the
following idea, which is sometimes referred to as Wyld's approach.
We introduce a random external force ${\bf f}$ in the Eq.~(1.1),
$$ \eqalign { \rho \left ({ \partial {\bf v} \over \partial t}
+ ({\bf v \nabla})
{\bf v} \right) &= - {\rm \bf grad}~p + \nu \Delta {\bf v} + {\bf f}\cr
   {\rm div}~{\bf v}   = 0} \eqno (1.6)$$
and we presume that the random force is Gaussian with the correlation function
$$ \langle {\bf f} ({\bf x},t) {\bf f}({\bf y}, t^\prime)
\rangle = {\bf D} ({\bf x}-{\bf y}) \delta (t-t^\prime) \eqno (1.7)$$
where ${\bf D}$ is some function\footnote{${}^{\dagger}$}
{\ninerm Note that here the average
is understood in the sense different from that of  Eq.~(1.4) because
it is not an equal-time correlation function, but rather a time dependent one.
Also, here we induce a randomness on the system with the random external
force, while in turbulence the randomness should be generated by the
equation itself.
Later throughout the paper we are going to study only time independent
equal time correlation
functions which are generated by  a very complicated motion of the
fluids. }.
 Then we solve Eq.~(1.6) perturbatively
in powers of nonlinear interaction term and apply the renormalization
group technique to interpret this solution as the renormalization of
viscosity and the coefficient in front of the interaction term (which
is 1 in Eq.~(1.6)). The hope is then to pass somehow to the limit when
the function ${\bf D}$ acts only at large distances thus reproducing the
behavior of the fluid in the turbulence conditions.
However, all the calculations performed in this fashion showed that in
this limit the equations of motion tend to the nontrivial fixed point
of renormalization group which cannot be treated perturbatively
or even worse, everything diverges to meaningless infinities$^{7}$.
So some other technique should be employed when studying turbulence.

In this paper, following mainly the reference [8], we shall try to
revive the old approach to turbulence which assumes the existence of
the stationary probability distribution. We shall study the wave
turbulence, a theory which, unlike hydrodynamics, can be solved
perturbatively, and we shall find the turbulent probability distribution
for it. We shall see that the turbulent probability distribution
is essentially different from the thermodynamics one, and the hope
is, some of the features we shall examine are independent of the
particular model choice and are present in any kind of turbulence.

\vglue 0.6cm
\leftline{\twelvebf 2. Probability Distributions}
\vglue 0.3cm
\leftline{\twelveit 2.1 The Turbulent Distributions}
\vglue 1pt
In this section we attempt to explain why Gibbs distribution cannot
describe the turbulent states of fluids or other statistical systems.
Let us review the derivation of Gibbs distribution. It usually starts with
claiming that any
stationary probability distribution should be of the form
$\exp (-F) $
where $F$ is the additive integral of motions, as follows
from the Liouville theorem$^3$. Then
they claim the only additive integral of motion the system
can have is the energy $E$, that is how Eq.~(1.2) is obtained. In our
case Eq.~(1.2) does not work as we already learnt, so something
must have gone wrong with the above reasoning.   What really had
gone wrong was there were other integrals of motion$^9$.
In fact, any dynamical system has as many integrals of
motion as the number of its degrees of freedom, at least locally in time.
To be more explicit, the
initial conditions expressed in terms of the changing variables
give us a formal integral of motion.

To be more precise, let us
imagine we have a hamiltonian system with the Hamiltonian $H(P,Q)$,
$P$ and $Q$ being the momenta and the coordinates.
The equations of motion are
$$ \eqalign { { dP \over dt} &= - { \partial H \over \partial Q } \cr
	      { dQ \over dt} &= {\partial H \over \partial P} }
\eqno (2.1.1) $$
One can show that the Eq.~(1.1) with the viscosity $\nu=0$
are also hamiltonian$^6$, so the motion of the incompressible
inviscid fluid is a particular case of the Eq.~(2.1.1).

Let us now imagine we somehow solved the Eq.~(2.1.1) and found
the coordinates and the momenta and the functions of time
and the initial values of the coordinates and the momenta, or
$$ \eqalign { P &= P(Q_0, P_0; t) \cr
	      Q &= Q(Q_0, P_0; t) } \eqno (2.1.2) $$
with
$$ \eqalign { P(Q_0, P_0; 0) &= P_0 \cr
	      Q(Q_0, P_0; 0) &= Q_0 } \eqno (2.1.3) $$
Then we can invert the above functions to get
$$ \eqalign { Q_0 &= Q_0(Q, P; t) \cr
	      P_0 &= P_0(Q, P; t) } \eqno (2.1.4) $$
The functions $Q_0$ and $P_0$ we thus obtained are the integrals of
motion {\twelveit by definition}. They explicitly depend on time, but
we can go on to solve for $t$ the first of the equations (2.1.4) and
substitute that to the second one. Or
alternatively, we can try to pass to the limit $t \rightarrow \infty$.

We can suspect a strange paradox here since the integrals of motion
(2.1.4) exist for any systems, even for nonintegrable ones which, by
definition, should
not possess too many integrals of motion! The resolution to this paradox
lies in the fact that, after the elimination of time, those integrals
of motion will become nonsinglevalued functions, depending on
the expressions like ${\rm arctan}(P/Q)$ which are called the angle
variables in mechanics.  Yet they
can perfectly be used to construct the probability distributions
other than thermodynamic equilibrium for the systems with infinite
numbers of degrees of freedom$^8$.

What is most exciting about those integrals of motion is that
some of them give us the turbulent
distributions, that is those distributions describe the systems
with fluxes of energy or other conserved quantities, while the rest
are anomalously not conserved.
To search for the right probability distribution from so many ones given
by Eq.~(2.1.4), we must solve the kinetic equation, which can be understood
as an anomaly cancellation condition. We shall discuss
it in the next section of this paper.

Of course, to construct those integrals for  Eq.~(1.1) is
a hopeless task. To do that, we would need to solve those equations
for any initial data, while it is clearly impossible to do.
Here we want to try to achieve a more modest aim, namely to find those
integrals for the turbulence of waves.

We end our discussion of turbulence on general terms at this point.
In what follows, we
are going to discuss only the turbulence of waves unless otherwise
stated.

\vglue 0.3cm
\leftline {\twelveit 2.2 Wave Turbulence}
\vglue 1pt
The turbulence of waves is the flux-states of the systems
consisting of waves with a small interaction$^6$. Its Hamiltonian
can be written down in the form
$$
H = \sum_{p} \omega_{p} a^\dagger_{p} a_{p} +
\sum_{p_1 p_2 p_3 p_4}
\lambda_{p_1 p_2 p_3 p_4} a^\dagger_{p_1} a^\dagger_{p_2} a_{p_3} a_{p_4}
\eqno (2.2.1)$$

It is just a collection of waves with the energy spectrum $\omega_{p}$ and
the four wave interaction $\lambda_{p_1 p_2 p_3 p_4}$ with the evident
properties $\lambda_{p_1 p_2 p_3 p_4} = \lambda_{p_2 p_1 p_3 p_4} =
\lambda_{p_3 p_4 p_1 p_2}$. $H$ is a classical Hamiltonian and $a_p$ are
classical variables while $a^\dagger_p$ are just their complex conjugates.
Let us note that this Hamiltonian conserves
the total wave number
$$ N=\sum_{p} a^\dagger_{p} a_{p}. \eqno (2.2.2) $$
We could also have considered the three wave interaction$^6$, but we are not
going to do it here.

The alluring property of this system is that the interaction
is small, which allows us to use the perturbation theory in all the
calculations. On the other hand, there are many real systems
existing in Nature which are described by Eq.~(2.2.1), the most notable
of them being the gravitational waves on the surface of water.

This system can be in the state of the thermodynamic equilibrium, which
would be described by the Gibbs distribution
$$ P= \exp (-{H + \mu N \over T}). \eqno (2.2.3)$$
But it could also exhibit the turbulent behavior. Let us imagine there
is a source of energy at some small value $p=a$ and a sink at
$p=\Lambda\gg a$.
Then there will be a flux of energy in between these two values of $p$.
(Actually there have to  be two sinks, to absorb both the energy and
the wave number, but it is irrelevant to the present discussion.)

To find the probability distributions which are more general
than Eq.~(2.2.3), we need to find the integrals of motion more general than
$H$ and $N$. And to do that,
we need to solve the equations of motion. It can be done perturbatively
in powers of the interaction. However, there is a simpler
way to achieve this aim.

Let us look for the integrals of motion in the form of the infinite series
$$ \eqalign {F=\sum_{p} f_p a^\dagger_p a_p +
\sum & \Lambda_{p_1 p_2 p_3 p_4}
a^\dagger_{p_1} a^\dagger_{p_2} a_{p_3} a_{p_4} +
\cr + & \sum \Omega_{p_1 p_2 p_3
p_4 p_5 p_6} a^\dagger_{p_1} a^\dagger_{p_2} a^\dagger_{p_3} a_{p_4} a_{p_5}
a_{p_6} + \dots.  } \eqno (2.2.4) $$
Here $\Lambda$, $\Omega$, etc.  are some still unknown functions.
We impose the
condition on $F$ that it is an integral of motion, or $\{ HF \} =0$, $\{$
and $\}$ being the Poisson brackets. With the help of the obvious
definition of the Poisson brackets,
 $$ \{a_p, a^\dagger_{p^\prime} \} =i \delta_{p p^\prime} $$
we  find those functions to be
$$ \Lambda_{p_1 p_2 p_3 p_4} = {f_{p_1} + f_{p_2} - f_{p_3} - f_{p_4} \over
\omega_{p_1} + \omega_{p_2} - \omega_{p_3} - \omega_{p_4}  }
\lambda_{p_1 p_2 p_3 p_4} \eqno (2.2.5)$$
$$ \Omega_{{p_1} {p_2} {p_3} {p_4} {p_5} {p_6}} = 4 \sum_{{p_7}}
{ ( \lambda_{{p_7} {p_1} {p_5} {p_6}} \Lambda_{{p_2} {p_3} {p_4} {p_7}} -
\Lambda_{{p_7} {p_1} {p_5} {p_6}} \lambda_{{p_2} {p_3} {p_4} {p_7}} )
\over \omega_{p_1} + \omega_{p_2} + \omega_{p_3}
- \omega_{p_4} - \omega_{p_5} -
\omega_{p_6} } , \eqno (2.2.6)$$
and so on. We can in principle
find recursively all the terms in the series (2.2.4), one after another.

However, in this form $F$ is not very well defined. The denominators of
$\Lambda_{p_{1}
p_{2} p_{3} p_{4}}$,
$\Omega_{p_{1} p_{2} p_{3} p_{4} P_{5} p_{6}}$ and, as one can check, of
all the higher order terms
in the series (2.2.4) may be equal to zero  for certain values of
of the indices $p_{1}$, $p_{2}$, etc. Then the sums in (2.2.4) are not
defined.

If we had the finite number of degrees of freedom, that would mean the
destruction of the integrals $F$. This phenomenon is well known
in mathematical physics$^{10}$, the destruction of the integrals
meaning that the system is no longer integrable after the interaction was
turned on.

However, here we assume that we have an infinite
number of $a_p$ and $a^\dagger_p$. Then the infinite number of degrees of
freedom means the sums in (2.2.4) turn into the integrals. We can shift
the poles in (2.2.5) and (2.2.6) to the complex plane to define
$$ \Lambda_{p_1 p_2 p_3 p_4} = {f_{p_1} + f_{p_2} - f_{p_3} - f_{p_4} \over
\omega_{p_1} + \omega_{p_2} - \omega_{p_3} - \omega_{p_4} - i \epsilon }
\lambda_{p_1 p_2 p_3 p_4} \eqno (2.2.7)$$
$$ \Omega_{{p_1} {p_2} {p_3} {p_4} {p_5} {p_6}} = 4 \sum_{{p_7}}
{ ( \lambda_{{p_7} {p_1} {p_5} {p_6}} \Lambda_{{p_2} {p_3} {p_4} {p_7}} -
\Lambda_{{p_7} {p_1} {p_5} {p_6}} \lambda_{{p_2} {p_3} {p_4} {p_7}} )
\over \omega_{p_1} + \omega_{p_2} + \omega_{p_3}
- \omega_{p_4} - \omega_{p_5} -
\omega_{p_6}- 2 i \epsilon } . \eqno (2.2.8)$$
where $\epsilon$ is a small constant.
Then the integrals the sums of Eq.~(2.2.4) turned into become well defined.
We should also shift the poles in all the higher terms of Eq.~(2.2.4).
Of course, we must pass to the limit $\epsilon \rightarrow 0$ in all the
calculations.

Let us note that we wrote $2 i \epsilon$ in the denominator
of Eq.~(2.2.8).
To understand why we should do so,
we must resort to the original way of finding
$F$ as explained in the previous subsection.
We start by assigning the variables $a_p$ and $a^\dagger_p$ the
initial values $a_{p}(t=0)=a^0_p$, $a^{\dagger}_p(t=0)=a^{\dagger 0}_p$.
Then by solving the equations of motion one can find $a(t)$ and $a^\dagger(t)$
as functions of $a^0$ and $a^{\dagger 0}$ and time. By inverting those
functions, one can find $a^0$ and $a^{\dagger 0}$ as functions of time,
$a(t)$ and $a^\dagger (t)$.
The quantity
$$ F=\sum_{p} f_p a^{\dagger 0}_{p} a^0_p, \eqno (2.2.9)$$
where $f_p$ are some arbitrary coefficients, is an additive integral of
motion by definition. Now if we try to solve the equations of motion
following from (2.2.1) perturbatively in powers of interaction, we shall
find that as soon as we express $F$ in terms of $a_p$ and $a^\dagger_p$
it turns into something close to Eq.~(2.2.4). The only difference will
be the explicit presense of time in form of the coefficients
${1 - \exp(- i \omega t) \over \omega}$. To obtain the time independent
integrals of motion, we shall pass to the limit $t \rightarrow \infty$
using the well known  formula
$$\lim_{t \rightarrow \infty} { 1 - \exp (-i \omega t) \over \omega}=
\lim_{\epsilon  \rightarrow 0} {1 \over \omega - i \epsilon}
\eqno (2.2.10)$$
As soon as we do that to Eq.~(2.2.9) it will turn itself into  (2.2.4).
This treatment shows that $\epsilon$ is actually an inverse time cutoff.

The pole shift to the complex plane in construction of the integrals
of motion is very important for the proper description of turbulence.
If one computes $F$ in the first two orders of
perturbation theory by solving the equations of motion and passing
to the limit $t \rightarrow \infty$,
one notices that the denominator of (2.2.6) is formed by
adding up the denominators of (2.2.5) and that is why we put
$2 i \epsilon$ in there. By induction, it is possible to prove
that we will get $n i \epsilon$ in the denominator of the $n$th term
of Eq.~(2.2.4). Writing $n i \epsilon$ is very essential for
consistency of some of the calculations$^8$, even though we shall
not encounter such calculations in this paper.

\vglue 0.3cm
\leftline{\twelveit 2.3. Turbulent Partition Function}
\vglue 1pt
The integrals of motion $F$ are the most general integrals the
system described by Eq.~(2.2.1) can have. By choosing different
functions $f_p$ we can obtain different integrals of motion including
the ones we knew before.

For example, if $f_p=1$, then Eq.~(2.2.4) turns into
$$ F=\sum_p a^\dagger_p a_p \eqno (2.3.1) $$
which is just the wave number quantity $N$.
If we choose $f_p=\omega_p$,
then $F$ will simply coincide with the Hamiltonian.
The corresponding probability distribution
$$ \exp(-F) \eqno (2.3.2)$$
will just be the Gibbs one.

However, if we choose different $f_p$ we shall obtain completely
new distributions. We shall see later that some of them describe the
turbulent states with constant fluxes while the rest will describe
nonstationary statistical states.

The turbulent partition function is given by
$$ Z= \int \prod_{p} da_{p} \prod_{p} da^\dagger_{p}
\exp( -F) \eqno (2.3.3) $$
and any correlation function can be computed using
$$ \langle X \rangle \equiv {1 \over Z}
\int \prod_{p} da_{p} \prod_{p} da^\dagger_{p} X \exp( -F) \eqno (2.3.4) $$

The turbulent partition function looks not too much different from the
standard statistical one. Nevertheless, it possesses many unusual properties.
First, the turbulent $F$ is an infinite series in powers of the interaction
and each next term is not independent of the previous ones, but
rather has a special form. In fact, only the first term of the series
(2.2.4) is arbitrary, all other terms are defined by the first one.
Secondly, we need to employ the pole shifting trick to define $F$
properly and we have to pass to the limit $\epsilon \rightarrow 0$
in the final answers. Nevertheless, one
can check by taking the complex conjugate of $F$ that it is
explicitly real so we shall not obtain unphysical complex values
for real correlation functions. Yet, the pole shifting will play
a very important role in the whole technique.

We would like now to turn to the discussion of what makes some
of $F$ describe the constant fluxes.

\vglue 0.6cm
\leftline {\twelvebf 3. Kinetic Equation}
\vglue 0.3cm
\leftline {\twelveit 3.1. Kinetic Equation in the first order of perturbation
theory}
\vglue 1pt
Gibbs distribution describes the thermodynamical states which do not
depend on time. One can check that explicitly with
$$\langle {dX \over dt}\rangle
=\langle\{HX\}\rangle=\int \{HX\} \exp (-H)=\int \{HH\} X
\exp(-H)\equiv 0 \eqno (3.1.1)$$
where we took the integral by parts and used that $\{H H\}\equiv 0$.
So, the average of derivative with respect to the time is always zero.
A really amazing fact is that the probability distribution $F$ can
give rise to the correlation functions which do depend on time.

Let us for example compute the average of the wave number change
with respect to the time, or
$$ \langle {d( a^\dagger_p a_p ) \over d t } \rangle \eqno
(3.1.2)$$
The equation expressing this quantity in terms of the functions $f_p$
and equating that with zero is called the kinetic equation.

By computing the Poisson bracket of the wave number with the Hamiltonian
we obtain
$$ \eqalign{\langle {d
(a^\dagger_p a_p) \over dt}\rangle &= \langle \{H, a^\dagger_p a_p\} \rangle =
 \cr
= - i \sum_{p_1 p_2 p_3 p_4}
\lambda_{p_1 p_2 p_3 p_4} & \langle a^\dagger_{p_1} a^\dagger_{p_2} a_{p_3}
a_{p_4} \rangle
(\delta_{p p_1} + \delta_{p p_2} - \delta_{p p_3} - \delta_{p p_4}) =0}
\eqno (3.1.3)$$

Since the function $F$ is real, one can easily see from (3.1.3) that only
the imaginary part of the correlation function $\langle a^\dagger_{p_1}
a^\dagger_{p_2} a_{p_3} a_{p_4} \rangle$ will contribute to the kinetic
equation
which can actually written down in the form
$$  - 4  i \sum_{p_2 p_3 p_4} \lambda_{p p_2 p_3 p_4}
{\rm Im} \langle a^\dagger_{p} a^\dagger_{p_2} a_{p_3}
a_{p_4}  \rangle = 0 \eqno (3.1.4) $$

Let us find the four point correlation function perturbatively. Its
zero order value is equal to
$$\eqalign {
\langle a^\dagger_{p_1}  a^\dagger_{p_2} a_{p_3} a_{p_4} \rangle
\equiv {1 \over Z} \int
\prod_{p} da_{p} & \prod_{p} da^\dagger_{p} a^\dagger_{p_1} a^\dagger_{p_2}
a_{p_3} a_{p_4} \exp( - \sum f_p a^\dagger_p a_p) = \cr
& ={1 \over f_{p_1} f_{p_2}}
(\delta_{p_1 p_3} \delta_{p_2 p_4} + \delta_{p_1 p_4} \delta_{p_2 p_3})}
. $$
This expression is real so it will not contribute to Eq.~(3.1.4).
Now the first order contribution to that function is obtained by
expanding the exponent in powers of $\lambda$ and is given by
$$ - 4\lambda_{p_1 p_2 p_3 p_4} {f_{p_3}+f_{p_4}-f_{p_1}-f_{p_2} \over
\omega_{p_3}+\omega_{p_4}-\omega_{p_1}-\omega_{p_2} - i \epsilon}
{1 \over f_{p_1} f_{p_2} f_{p_3} f_{p_4}}  \eqno (3.1.5) $$

Imaginary part of Eq.~(3.1.5) is
$$  4 i \pi \lambda_{p_1 p_2 p_3 p_4} (f_{p_1} + f_{p_2}
- f_{p_3} - f_{p_4}) \delta (\omega_{p_1} + \omega_{p_2} - \omega_{p_3} -
\omega_{p_4}) {1 \over f_{p_1} f_{p_2} f_{p_3} f_{p_4} }$$
and after substituting it to (3.1.4) we arrive at the kinetic
equation in the first order of perturbation theory
$$
16 \pi \sum_{p_2 p_3 p_4} \lambda^2_{p p_2 p_3 p_4} \delta (\omega_p -
\omega_{p_2} - \omega_{p_3} - \omega_{p_4}){ f_p+ f_{p_2}-f_{p_3}-f_{p_4}
\over f_{p} f_{p_2} f_{p_3} f_{p_4} } = 0 .  $$
or, taking into account the infinite number of degrees of freedom,
$$
16 \pi \int {dp_2 dp_3 dp_4} \lambda^2_{p p_2 p_3 p_4} \delta (\omega_p -
\omega_{p_2} - \omega_{p_3} - \omega_{p_4}){ f_p+ f_{p_2}-f_{p_3}-f_{p_4}
\over f_{p} f_{p_2} f_{p_3} f_{p_4} } = 0 .  \eqno (3.1.6)$$

The Eq.~(3.1.4) or its first order approximation Eq.~(3.1.6) is of special
importance.  The probability distributions $F$ do not always give the
stationary states of the systems we study. Only when we choose
$f_p$ such that Eq.~(3.1.4) is satisfied, the corresponding $F$ is a true
turbulent distribution. So, to solve the turbulence of waves means two things.
One is to solve the kinetic equation and find such $f_p$ that it is
satisfied. The second is to find the correlation functions using
$F$ with those $f_p$ we found.

Let us note that the thermodynamic equilibrium provides the simplest
solution because the four point correlation function used in Eq.~(3.1.4)
is always real when computed with $F\equiv H$ or $F \equiv N$. Really,
the only place where complexity can come into play is the term $i \epsilon$
and it appears neither in $H$ nor in $N$. In the language of the first
order equation Eq.~(3.1.6) $f_p=1$ or $f_p=\omega_p$ makes the
integrand of Eq.~(3.1.6) to be zero. The integrand has an evident
physical meaning of the wave exchange between fixed wave numbers.

However,
there are other solutions of Eq.~(3.1.4) which give us nonzero fluxes.
They can be found explicitly in the first order of perturbation theory
by solving the Eq.~(3.1.6)$^6$. They occur when the integrand of Eq.(3.1.6)
is nonzero while the integral is zero, providing the equilibrium without
the detailed balance.
As soon as we solve the Eq.~(3.1.4) and find uniquely the function $f_p$ and
therefore the function $F$ as well, we can start computing correlation
functions and the problem of turbulence will be at least formally solved.

Now as is well known in the phase transition theory, the first order
approximation does not always give the right answer. The
first order approximation
 is called mean field theory in the
physics of phase transitions and is called {\twelveit weak turbulence} in
our theory.  However, there are important cases where we must sum up all
orders of perturbation theory$^8$. They are called {\twelveit strong
turbulence}.  Weak turbulence has been well studied in the last two
decades.  Strong turbulence is much more difficult to study. It is
especially alluring to study the strong case because the original problem of
turbulent fluid does not have any small parameter and as such is
the problem of strong turbulence.

We shall postpone further discussion of weak and strong turbulence
till next section.

\vglue 0.3cm
\leftline{\twelveit 3.2. The Anomalous Nature of Kinetic Equations}
\vglue 1pt
We saw in the previous subsection that the distributions $F$ can
in general describe nonstationary statistical states. That can
easily lead to the apparent paradox. We can repeat the line (3.1.1)
for the distribution $F$.
$$\langle {dX \over dt}\rangle=\langle\{HX\}\rangle=
\int \{HX\} \exp (-F)=\int
\{HF\} X \exp (-F)\equiv 0 $$
because $\{HF\}=0$.
But that clearly contradicts the derivation of the kinetic equation.
We saw that the average
$$ \langle { d( a^\dagger_p a_p) \over d t} \rangle$$
is not equal to zero apart from the specially chosen $f_p$.

The resolution of this paradox lies in the fact that $F$ is
the integral of motion only in the limit $\epsilon \rightarrow 0$. If we
compute $\{H F\}$ we shall  discover that it is of the order of $\epsilon$.
But then, when computing the integral $\int \{H F\} X \exp (-F)$, we shall
find that it in itself is of the order of $1/ \epsilon$.  Those epsilons
cancel and we can safely pass to the limit $\epsilon \rightarrow 0$ to get
the finite answer for the kinetic equation.  The effect is completely
analogous to the anomaly of Quantum Field Theory. By imposing the cutoff
$\epsilon$ on the theory we violate the conservation of
$F$ and it  is never
fully restored.

The kinetic equation can thus be interpreted as a condition for
anomaly cancellation. As soon as it is fulfilled, $F$ becomes a true
integral of motion, even within this regularization scheme.

The anomaly shows how important those $\epsilon$'s are for the theory.
While the distribution $F$ we found works only for wave turbulence,
that may be a general feature of all the turbulent distributions
including those for Eq.~(1.1).

\vglue 0.6cm
\leftline {\twelvebf 4. Turbulent Field Theory}
\vglue 0.3cm
\leftline {\twelveit 4.1. Diagrammatic technique}
\vglue 1pt
To be able to go beyond the first order of perturbation theory,
we must learn how to compute correlation functions in arbitrary
order. The function $F$ contains arbitrary
high powers of $\lambda$ and it is not so easy to expand $\exp(-F)$.
However that can be done in general.  It is possible
to formulate the result of the expansion in terms of relatively
simple Feynman rules, though more complex than the standard ones$^8$.

$F$ contains interactions of the fourth, fifth and all the higher
orders, but it turns out it is enough to use the four-vertices
in the diagrams because the higher order terms like (2.2.6) are
effectively expressed in terms of the four point one.

Since the quadratic term in $F$ has the form $-f_p a^\dagger_p a_p$,
the propagator has the form $1/f_p$. The interaction coefficient
is $\lambda_{p_1 p_2 p_3 p_4}$, it will ``live'' at each vertex of
the diagram. The most nontrivial part is a prefactor coming from a
very complex form of the interaction. The rules of computing that
prefactor are not too simple and are given in reference [8].
Here we shall give that expression just for one diagram to give
a general idea of what should be expected.

Let us consider, as an example, the simplest bubble diagram one can
think of in the $\varphi^4$-like theory. Let us call the momenta of
its external legs
as 1, 2, 3, and 4, and  of its two intermediate lines as 5 and 6.
Its expression will be
$$ \eqalign { & {\lambda_{1 2 3 4} \lambda_{1 2 5 6} \lambda_{5 6 3 4} \over
f_1 f_2 f_3 f_4 f_5 f_6} \times \cr  \times \biggl( & {(f_1+f_2)(f_5+f_6)
\over (\omega_1+\omega_2-\omega_3-\omega_4-2 i \epsilon)
(\omega_5+\omega_6-\omega_3-\omega_4 -i \epsilon)} + \cr + & {(f_3+f_4)
(f_5+f_6) \over (\omega_1+\omega_2-\omega_3-\omega_4-2 i \epsilon)
(\omega_1+\omega_2-\omega_5-\omega_6-i \epsilon)}
 - \cr - &{(f_1+f_2)(f_3+f_4)
 \over
(\omega_1+\omega_2-\omega_5-\omega_6-i
\epsilon)
(\omega_5+\omega_6-\omega_3-\omega_4-i \epsilon)} \biggl) } \eqno (4.1.1)$$
The expression begins with the standard product of propagators
and vertices. But then we see a prefactor which is completely
different from what was encountered in Field Theory before.
It was the higher order vertices generated by $F$
which combined in the nontrivial way
to give rise to that prefactor. While it looks a little horrible, it does
have at least one nice feature though, it goes to 1 if $f_p \rightarrow
\omega_p$ or
if the turbulent distribution goes into the thermodynamic one.
The ``turbulent" field theory we obtained can thus be considered a
direct generalization of the standard Field Theory techniques.

Let us review the differences of the turbulent field theory
from the standard one. Its action is a nonlocal expression given by
an infinite series given by Eq.~(2.2.4). There is an additional
$\epsilon$-regularization so in effect the turbulent action is a
generalized function. The correlation functions may depend on time
and the cancellation of that dependence gives us additional condition,
called the kinetic equation, to be satisfied for the action to describe
the turbulence. And the diagrams obtained from the perturbative expansion
differ from standard ones by a special prefactor.

This theory has not been studied well yet.
One may  ask what ``classical'' equations of motion its action
corresponds to, if many notions of field theory like the existence
of Hilbert space of states and others can be applied to it.
And of course whether it can be
 constructed in cases what the perturbation theory does not work.
 These are the questions to be answered in future.

\vglue 0.3cm
\leftline{\twelveit 4.2. $\epsilon$-expansion}
\vglue 1pt
Now it is time to discuss how to actually compute something meaningful
in this theory.
The program is clear. For a given $\omega_p$ and $\lambda_{p_1 p_2 p_3 p_4}$
we must solve the equation (2.1.4) for $f_p$ and then, for the $f_p$
found we must compute the correlation functions which may be of interest.
The most interesting of them is the wave spectrum
$$ n_p=\langle a^\dagger_p a_p \rangle    \eqno (4.2.1) $$

To find the solutions of the equation (3.1.4) or even of its first order
approximation (3.1.6) which are different from the thermodynamic
equilibrium is a very difficult task.
The only approachable case is when the spectrum $\omega_p$ and
the interaction $\lambda$ are scale invariant
$$ \omega_p=p^\alpha,~\lambda_{p_1 p_2 p_3 p_4}=(p_1 p_2 p_3 p_4)^{\beta \over
4} U \delta ({\vec p_1} + {\vec p_2} - {\vec p_3} - {\vec p_4}) \eqno (4.2.2)
 $$
where U depends only on the ratio of the momenta and their mutual angles.
Then we can apply the Zakharov's transformations$^6$ to solve (3.1.6) to
get
$$ f_p=p^{\gamma}, ~\gamma={2 \over 3} \beta + d {\rm \ or \ } \gamma=
{2 \over 3} \beta + d - {\alpha \over 3} \eqno (4.2.3) $$
where $d$ is the number of space dimensions. The first of the
solutions corresponds to the energy flux while the second one to the
flux of particle number.

While the exact derivation of Eq.~(4.2.3) is rather complicated, there
is a nice way of interpreting it. Let us take the Eq.~(3.1.6) and
make a change of variables in the integral
$$p_2=x_2 p,~p_3=x_3p,~p_4=x_4p \eqno (4.2.4) $$

We shall get
$$ {\partial n_p \over \partial t}=p^{- 3 \gamma  + 2 \beta +3 d - 1 -
 \alpha} 16 \pi  \int {dx_2 dx_3 dx_4}
K(x_2, x_3, x_4) ~\delta (1 + x_2^\alpha - x_3^\alpha - x_4^\alpha){ 1 +
x_2^\gamma-x_3^\gamma-x_4^\gamma \over (x_2 x_3 x_4)^\gamma } = 0 .  \eqno
(4.2.5)$$
where $K$ is
$$ K(x_2, x_3, x_4) = p^{- 2 \beta +d }
\int do_2 do_3 do_4 \lambda^2_{p p_2 p_3 p_4}, $$
$do_2$, $do_3$, and $do_4$ are the angle integrations associated with
$d^d p_2$, $d^d p_3$, and $d^d p_4$ respectively. It may  not be evident
at  a first glance, but
 by construction, $K$ does not depend on $p$.

Let us compare Eq.~(4.2.5) with the continuity equation
$$ {\partial n_p \over \partial t} + {\partial J_p \over \partial p} =0
\eqno (4.2.6) $$
We see that the wave flux $J_p$ is
$$ J_p=-
p^{- 3 \gamma  + 2 \beta +3 d  -
 \alpha}  { X \over - 3 \gamma  + 2 \beta +3 d  -
 \alpha}  \eqno (4.2.7) $$
We denoted the integral in Eq.~(4.2.5) by $X$. This integral is just
a
number which is not dependent on $p$. The condition $J_p=$const (or
$\omega_p J_p$=const for the energy spectrum $\omega_p n_p$) coincides with
(4.2.3). The denominator of Eq.~(4.2.7) is then equal to 0, but so is $X$.

However, (4.2.3) must be regarded as just the first approximation
to the answer, just as mean field theory is just the first approximation
for the critical phenomenon correlation functions and they often
give the wrong critical indices$^5$.
The criterion for the applicability of the mean field theory
techniques is, as always, the dimension of the interaction constant
$\lambda_0$. We can easily find it by analyzing $F$ with $f_p=p^{- \gamma}$
as a zero approximation. We obtain
$$ \kappa=-{\rm dim}(\lambda_0)= \beta+ d-\gamma-\alpha \eqno (4.2.8)$$
or, after substituting Eq.~(4.2.3),
$$ \kappa={\beta  \over 3} - \alpha {\ \ \rm or }~ \kappa={\beta  \over 3}
- {2 \over 3} \alpha \eqno (4.2.9)$$

The most striking difference between this     parameter  and
the $\epsilon$-parameter of the phase transition theory is that
it does not depend on the number of space dimensions at all.
Also, the convergence or divergence of the integrals we encounter in
the computations is governed not only by the power counting but also
by the parameter $U$ of the interaction in Eq.~(4.2.2). We may always
choose $U$ so as to make all integrals to be always convergent, for
example, by taking $U$ to be exponentially decaying function at large
ratio of momenta.

However, recently there was an attempt by the author$^8$
to study the $\epsilon$-expansion
for the case when $U$=1. Then the interaction does not influence
the convergence or divergence of the integrals at all. If $\kappa >0$ then
mean field theory is exact (weak turbulence),
while if $\kappa < 0$ (strong turbulence), then it had been shown that
taking into account higher diagrams computed according to the rules mentioned
above
we can compute a correction to the mean-field
spectrum $\gamma$ in the form
$$ \gamma \rightarrow \gamma - {2 \over 3} \kappa \eqno (4.2.10)$$
when $\kappa$ is a small negative number and $\alpha < 1$.

\vglue 0.6cm
\leftline{\twelvebf 5. Open problems}
\vglue 0.4cm
We have studied in this paper the probability distribution for the
turbulent states of the statistical systems. While we did that for the
turbulence of waves, we hope that some of the distinctive features
of the distribution we discovered are universal.

Let us return now to the discussion of the main task
of  solving  the original problem  of the
hydrodynamics turbulence.

We saw that the Hamiltonian formulation was rather important for the
discussion of turbulence. There exists a Hamiltonian formulation of
hydrodynamics, the so-called Clebsch variables. We define
$$ {\bf v} = {\rm \bf grad}~\varphi + \lambda {\rm \bf grad}~\mu
\eqno (5.1)$$
The field $\varphi$ is a dependent variable due to the incompressibility
equation ${\rm div}~{\bf v}=0$. Then it turns out the variables $\lambda$ and
$\mu$ are the canonical ones, with the hamiltonian given by
Eq.~(1.3) where we substitute Eq.~(5.1) for the velocity.
It can be rewritten in the momentum space as
$$ H= \int_{p_1 p_2 p_3 p_4} \lambda_{p_1 p_2 p_3 p_4} a^\dagger_{p_1}
a^\dagger_{p_2} a_{p_3} a_{p_4} \eqno (5.2) $$
where $a$ and $a^\dagger$ are the Fourier transforms of $\lambda+i \mu$
and $\lambda- i \mu$ and $\lambda_{p_1 p_2 p_3 p_4}$ are
 a specifically
chosen the interaction coefficient (see reference [6] for the derivation
of the fluid motion reformulation in terms of the canonical variables)

We see that the Hamiltonian (5.2) looks like the wave Hamiltonian
given by Eq.~(2.2.1) without the wave part, $\omega_p=0$. One can
ask if we can somehow interpret the turbulent field theory we
discussed above in the limit $\omega \rightarrow 0$. The staightforward
limit is impossible because each higher order diagram will contain
a higher power of $\omega$ in its denominator. Another method
is to introduce  ``mean field'' frequencies$^{11}$ by
$$\omega_p \propto \int_{q} dq~ \lambda_{p q p q} \langle {a_q a^\dagger_q}
\rangle. $$ Not surprisingly, this immediately leads to the
Kolmogorov spectrum as an answer, as any mean field theory should do.

The only hope of honestly taking that limit,
we believe, lies in conformal field theory.
In two dimensions the turbulent field theory gives a set of  the
scale invariant correlation functions and it is possible they are
conformally invariant as well. That may be directly checked, however,
in case of the turbulence of waves,  which would be very interesting
to do and which has not been done yet.

Another important task here is to understand better the behavior
of the infrared modes. The turbulent correlation functions
grow with distance.    Let us take for example the function
Eq.~(1.5).
Of course,  its validity  is something to be proved, but
yet the future full theory will probably give just a small correction
to the index $-{5 \over 3}$ of Eq.~(1.4).

The Eq.~(1.5) shows there is a strong infrared divergence of any
integrals involving the velocity correlation functions, and that
means everything depends on what is going on at the boundary.
Those divergencies prevented a complete construction of conformal
theory of two dimensional fluid turbulence$^9$.

A similar divergency occurs in the wave turbulence theory discussed
in this paper. There $\gamma > d$ and because of that there is an
infrared divergency there as well.
One can actually see that this IR divergency appears in many of the
integrals in the perturbative calculations. There has not been,
to the author's knowledge, any attempts to interpret them in one
way or another.

A nice way of thinking of  those divergencies is by assuming the
infrared modes of the theory form some kind of a condensate, like
the Bose condensate of the weakly interacting Bose gas. The overall
infrared motion of the waves acts as a background for the small
fluctuations around it. This picture
can be checked quantitatively in the theory of wave turbulence. The idea
would be to fix the amplitude of the zero mode $a_0=A$ like
in Bose gas condensation theory, substitute it into
Eq.~(2.2.4), and see whether there is a choice of $A$ which cancels
the IR divergencies.  It is still unknown if this or a similar computation
can provide an explanation of the behavior of zero modes.
However we do know from numerical experiments that large
scale motion is very important for the hydrodynamic turbulence.
In two dimensions we observe a few large vertices of fluid motion
and the turbulence is a fluctuation around those vertices. In three
dimensions we observe a condensate of vertex filaments. A future
theory of turbulence must be able to account for those effects
and it could be its formulation depends crucially on taking them
into account correctly.

The last thing we would like to mention  is the absence of any
nonperturbative exactly
solvable models in turbulence which is very frustrating.
All theories solved here exactly thus far had been trivial.
 On the other hand, in field
theory and statistical mechanics
a large number of nontrivial exactly
solvable models has been discovered
which by far increased our understanding of
the physics of field theory and their study is one of the most important
activities now in this field.
One may hope that soon the theory of
turbulence will also benefit from the discovery of such models.

\vglue 0.6cm
\leftline{\twelvebf 5. Acknowledgements}
\vglue 0.4cm
I would like to thank I. Kogan and A. Polyakov for many helpful discussions
on the subject presented in this paper. I am also grateful to the organizers
of the conference, especially to H. Saleur, for the
kind invitation to give a talk.

\vglue 0.6cm
\leftline{\twelvebf 6. References}
\vglue 0.4cm

\medskip
\itemitem{1.} L.D. Landau, E.M. Lifshits,
{\twelveit Fluid Mechanics} (Pergamon Press, 1980).
\itemitem{2.} A.S. Monin, A.M. Yaglom, {\twelveit Statistical Fluid
Mechanics:
Mechanics of Turbulence} (The MIT Press, Cambridge, 1971).
\itemitem{3.} L.D. Landau, E.M. Lifshits,
{\twelveit Statistical Physics, Part I} (Pergamon Press, 1980).
\itemitem{4.} P.B. Weichman, in {\twelveit Nonlinear Waves and Weak
Turbulence}, eds. N. Fitzmaurice, D. Gurarie, F. McCaughan,
W.A. Woyczynski  (Birkhauser, Boston, 1993),
p. 239.
\itemitem{5.} A. Patashinski and V. Pokrovski, {\twelveit Fluctuation Theory
of Phase Transitions} (Pergamon Press, Oxford, 1979).
\itemitem{6.}  V. Zakharov, V. L'vov and G. Falkovich,
 {\twelveit Kolmogorov spectra of turbulence I}
(Springer-Verlag Berlin Heidelberg, 1992).
\itemitem{7.} C. DeDominicis, P.C. Martin, {\twelveit Phys. Rev.} {\twelvebf
 A19} (1979) 419.
\itemitem{8.} V. Gurarie, Princeton preprint PUPT-1492, bulletin board
hep-th/9405077.
\itemitem{9.} A. Polyakov, {\twelveit Nucl. Phys.} {\twelvebf B396} (1993) 367.
\itemitem{10.} V.I. Arnold, {\twelveit Mathematical Methods of Classical
Mechanics} (Springer-Verlag, 1989).
\itemitem{11.} V. Zakharov,
in {\twelveit Nonlinear Waves and Weak
Turbulence},  eds. N. Fitzmaurice, D. Gurarie, F. McCaughan,
W.A. Woyczynski (Birkhauser, Boston, 1993),
p. 13.

\bye